\documentclass[a4paper]{article}

\usepackage{INTERSPEECH2021}
\usepackage{flushend}
\usepackage{cite}
\usepackage{enumitem,kantlipsum}
\DeclareMathOperator{\diag}{diag}

\newcommand{\mb}[1]{\mathbf{#1}}
\newlength\figureheight 
\newlength\figurewidth 

\title{Speech Decomposition based on a Hybrid Speech Model and Optimal Segmentation}
\name{Alfredo Esquivel Jaramillo, Jesper Kjær Nielsen, Mads Græsbøll Christensen}
\address{
  Audio Analysis Lab, CREATE, Aalborg University, Denmark}
\email{\{aeja,jkn,mgc\}@create.aau.dk\\}

\begin{document}

\maketitle
\begin{abstract}
In a hybrid speech model, both voiced and unvoiced components can coexist in a segment. Often, the voiced speech is regarded as the deterministic component, and the unvoiced speech and additive noise are the stochastic components. Typically, the speech signal is considered stationary within fixed segments of 20-40 ms, but the degree of stationarity varies over time. For decomposing noisy speech into its voiced and unvoiced components, a fixed segmentation may be too crude, and we here propose to adapt the segment length according to the signal local characteristics. The segmentation relies on parameter estimates of a hybrid speech model and the maximum a posteriori (MAP) and log-likelihood criteria as rules for model selection among the possible segment lengths, for voiced and unvoiced speech, respectively. Given the optimal segmentation markers and the estimated statistics, both components are estimated using linear filtering. A codebook-based approach differentiates between unvoiced speech and noise. A better extraction of the components is possible by taking into account the adaptive segmentation, compared to a fixed one. Also, a lower distortion for voiced speech and higher segSNR for both components is possible, as compared to other decomposition methods. 
\end{abstract}

\section{Introduction} \vspace{-2pt}

The problem of decomposing speech into its voiced and unvoiced components is useful in applications such as speech coding, analysis, synthesis, modification and diagnosing of illnesses \cite{elie2016robust, griffin1988multiband, aej2018, mehta2005synthesis, jackson2000characterisation, stylianou1995high, belalcazar2013automatic}. As implied by hybrid speech models (e.g., harmonic plus noise model) \cite{stylianou2004modeling,stylianou1995high}, deterministic and stochastic components may coexist in a speech segment. The deterministic part corresponds to voiced speech, which is well represented by a sum of harmonically related sinusoids \cite{jensen2001speech, MadsPitch, FastNLS, QUINN2021107860}, whose frequencies are an integer multiple of the pitch. The stochastic parts cover what is not described by the harmonic model, including for example, glottal turbulences and friction. The traditional speech decomposition methods \cite{Yegnanarayana, elie2016robust}, however, do not distinguish between the unvoiced speech and the additive noise, and this distinction may be relevant, e.g., in remote voice assessment applications \cite{moran2006telephony}. The method in \cite{elie2016robust} considers the colored nature of the stochastic parts of speech in order to estimate the pitch using fixed window lengths. However, the authors hypothesize that the decomposition performance can be improved by using adaptive windows instead and by estimating the number of harmonics. The iterative decomposition method in \cite{Yegnanarayana} is based on the cepstrum to obtain pitch information, which is not very robust under high noise conditions \cite{robust2020}. Moreover, in \cite{jackson2000characterisation} it was found to converge to the wrong solution. Often, a speech signal is assumed to be stationary within segments of a fixed length which last between 20 and 40 ms \cite{preference}. However, due to its non-stationary nature, the speech signal characteristics might change quickly during short periods of time \cite{drepper2005two}. Therefore, the optimal choice should be a time-varying segment length which better accommodates the local characteristics and has a better fit to a specified model. For example, if the pitch remains nearly constant, the segment length should be longer than when it exhibits fast variations \cite{ColoredWhite}. 

An effort based on linear filtering to estimate separately the voiced and unvoiced parts from noisy speech was presented in \cite{aej2018}. In this paper, instead of relying on conventional noise tracking methods (e.g., \cite{Martin2001}), the noise statistics were estimated using approaches which rely on prior spectral information contained in codebooks \cite{srinivasan2005codebook,eusipco2019}. In order to have a better recovery of the components, it is here proposed to do the extraction based on optimal segmentation \cite{ColoredWhite, RDprediction}, instead of using a fixed one. To find the best possible segmentation, the parameters of the deterministic and stochastic parts are iteratively estimated as described in \cite{robust2020} for the different candidate segments and candidate models. Such approach is more robust to high noise levels than classical pitch estimators \cite{robust2020}. Estimates of the unvoiced and noise AR parameters are obtained from a codebook-based procedure \cite{srinivasan2005codebook} which is able to assign a zero excitation variance in silent segments for the speech part if necessary. The parameter estimates on the optimal segments are used to apply linear filtering to yield estimates of the individual components. 


\section{Signal model and filtering for speech decomposition} \vspace{-2pt}

In this section, we describe how noisy speech can be decomposed into its components using linear filtering which require the knowledge of the different statistics. A speech segment of length $N$ is described with a hybrid speech model. The model assumes that for a clean speech signal \vspace{-6pt}
\begin{equation}
s(n)=v(n)+u(n),
\end{equation}
where the unvoiced part $u(n)$ is represented as an AR process and the voiced part $v(n)$ is described by the harmonic model. In speech decomposition, the goal is to extract both $v(n)$ and $u(n)$ when $s(n)$ is degraded by additive colored noise $c(n)$, i.e., \vspace{-6pt}
\begin{equation}
 y(n)=v(n)+u(n)+c(n).   
\end{equation} 
The additive noise is also modelled as an AR process. The observation $y(n)$ can also be expressed as $y(n)=v(n)+x(n)$, where $x(n)=u(n)+c(n)$ is the residual containing the stochastic parts of noisy speech. By considering a vector of $M$ ($<N$) samples $\mb y = \mb v + \mb u +\mb c$, where $\mb v$, $\mb u$ and $\mb c$ are assumed to be uncorrelated, the $M \times M$ covariance matrix of the observation is expressed as the sum of covariance matrices of each component, i.e. $\mathbf{R}_{\mathbf{y}} = E \left[  \mathbf{y}\mathbf{y}^T \right] = \mathbf{R}_{\mathbf{v}}+\mathbf{R}_{\mathbf{u}}+\mathbf{R}_{\mathbf{c}}= \mathbf{R}_{\mathbf{v}}+\mathbf{R}_{\mathbf{x}}$. Here, $E  \left[\cdot\right] $ denotes expectation and $(\cdot)^T$ denotes the transpose. 

Initially, we want to extract an estimate of the desired voiced speech vector $\mb v$, by applying a linear filtering $M \times M$ matrix to $\mb y$, i.e., $\hat{\mb v}=\mathbf{H}_{v} \mb y = \mathbf{H}_{v} \mb v+\mathbf{H}_{v} \mb x$, where \vspace{-4pt}
\begin{equation}
\mb H_v = [ \mb{h}_{v,1}^H \ \mb{h}_{v,2}^H \ \hdots \ \mb{h}_{v,M}^H ]^T,    
\end{equation}
$\mb{h}_{v,m}, m= 1,2,...,M$ are complex valued filters of length $M$ and $(\cdot)^H$ is the conjugate transpose. The filtering applied in the time domain is commonly used when voiced speech parts described by the harmonic model are considered \cite{norholm2015enhancement}. Several filter designs are possible from a recently introduced variable span linear filtering framework (VSLF) \cite{benesty2016signal}, by choosing a number of eigenvectors and eigenvalues of the joint diagonalization of $\mathbf{R}_{\mathbf{v}}$ and $\mathbf{R}_{\mathbf{x}}$.
We here use the $M$ eigenvectors $\mathbf{b}_q$ and eigenvalues $\lambda_q$ to form an $M \times M$ Wiener filtering matrix \cite{benesty2016signal}\vspace{-6pt}
\begin{equation} \label{voicFilter}
    \mathbf{H}_{v}=\mathbf{R}_{v}\sum_{q=1}^{M}\frac{\mathbf{b}_q\mathbf{b}_q^H}{1+\lambda_q}. \vspace{-2.5pt}
\end{equation}
In order to consider prior spectral information stored in codebooks for estimating $u(n)$, we make use of the corresponding representation of $\mathbf{R}_{x}$ in the frequency domain, i.e., the power spectral density (PSD) $\Phi_x(\omega) = \Phi_u(\omega)+\Phi_c(\omega)$, where $\Phi_u(\omega)$ is the unvoiced component PSD and $\Phi_c(\omega)$ is the noise PSD. Estimates of these PSDs can then be used to apply a frequency domain Wiener filter $H_u(\omega)=\frac{\hat{\Phi}_u(\omega)}{\hat{\Phi}_u(\omega)+\hat{\Phi}_c(\omega)}$ to yield an estimate of the unvoiced component $U(\omega)=H_u(\omega)\hat{X}(\omega)$, where $\hat{X}(\omega)$ is the spectrum of the residual which is obtained as described in the next section. 

\section{Statistics and parameters estimation} \vspace{-2pt}

We now describe how to estimate the required statistics and parameters in order to apply previously described linear filtering. The harmonic model of voiced speech assumes that this component is represented as a set of sinusoids having frequencies which are an integer multiple of the pitch $f_0$ \cite{MadsPitch, jensen2001speech}, i.e., \vspace{-5pt}
\begin{equation}
	v(n)=\sum_{l=1}^{L}\left[ \alpha_le^{j2\pi f_0ln}+\alpha_l^*e^{-j2\pi f_0ln}\right] , \label{harmModel2}
\end{equation}
for a segment of length $N$. This model, however, will have a more accurate fit for a particular segment length $N_{\text{opt}}$, which will be known after an optimal segmentation of the signal has been obtained. Here, $L$ is the unknown number of harmonics, $\alpha_l=\frac{A_l}{2}e^{j\psi l}$ is the complex amplitude of the $l$'th harmonic with $A_l>0$ the real amplitude, $\psi_l$ the initial phase and $^*$ the complex conjugate. A voiced vector of $M$ succesive samples can be written as
$\mb{v}= \mb Z (f_0)\bm{\alpha}$, where $\mb Z$ is a matrix of Fourier vectors, i.e., \vspace{-4pt}
\begin{align}
\mb Z (f_0)&= [\mb z(f_0) \, \,\mb z^*(f_0) \, \, \hdots \, \, \mb z(Lf_0)  \, \,  \mb z^*(L f_0) ], \label{eq:Z}\\
\mb z(l f_0) &= [1 \, \, e^{jl2\pi f_0} \, \, \hdots \, \, e^{jl2\pi f_0 (M-1)}]^T, \label{eq:z}
\end{align}
and $\bm{\alpha} = \frac{1}{2} [A_1e^{j\psi_1} \, \, A_1e^{-j\psi_1} \, \, \hdots \, \, A_Le^{j\psi_L} \, \, A_Le^{-j\psi_L}]^T$ is a vector containing the amplitudes of the harmonics. The unvoiced parts of speech are modelled as an AR process of order $P$ (often, set to a fixed value \cite{stylianou1995high,stylianou2004modeling}), i.e., \vspace{-5.5pt}
\begin{equation} \label{unvModel}
    u(n)=-\sum_{i=1}^{P}\beta_{u_i}u(n-i)+e(n), 
\end{equation}
where $\{\beta_{u_i}\}_{i=1}^P$ are the $P$ AR coefficients of the unvoiced speech and $e(n)$ is the excitation WGN process with variance $\sigma_e^2$. Similarly, the colored noise $c(n)$ is modelled as an AR process with the $P$ AR coefficients $\{\gamma_{c_i}\}_{i=1}^P$.  

The estimated voiced part covariance matrix $\mathbf{R}_{\mathbf{v}} = E \left[  \mb{v}{\mb{v}}^T \right]$ can be expressed as $\hat{\mathbf{R}}_{\mathbf{v}}=\mathbf{Z}(\hat{f_0})\hat{\mathbf{P}}\mathbf{Z}(\hat{f_0})^H$ \cite{stoica2005spectral}, where the estimated amplitude covariance matrix  has the form $
\hat{\mathbf{P}}=E\{\hat{\bm{\alpha}}\hat{\bm{\alpha}}^H\}=\frac{1}{4}\diag([\hat{A}_1^2 \, \hat{A}_1^2 \hdots \, \hat{A}_L^2 \, \hat{A}_L^2])$. 
It is therefore required to have estimates of the pitch $f_0$ and of the linear parameters. At a first instance, we would need to have estimates of these parameters from the optimal segment length $N_{\text{opt}}$ when we do the processing based on the optimal segmentation. However, to estimate $N_{\text{opt}}$, we first need to estimate the parameters for all the possible segment lengths. The optimal segment length maximises the a posteriori probability of the observed data \cite{ColoredWhite}, as described in the next section. First, based on the estimated noise statistics (e.g., \cite{Martin2001, eusipco2019}), a pre-processor is applied in order to pre-whiten the noise component \cite{eusipco2019}, yielding the pre-whitened signal $y_{\text{W}}(n)$. This will allow to have better pitch estimates (i.e., reduce the subharmonic errors) from the nonlinear least-squares (NLS) estimator based on WGN assumption \cite{FastNLS}. The parameters inside each possible candidate segment length are  estimated by an approximate joint estimator of the voiced speech and the stochastic parts parameters, by iterating between these two steps: \cite{robust2020}
\begin{enumerate} [wide, labelwidth=!, labelindent=0pt]
\item \noindent The $f_0$ is obtained from the NLS estimator \cite{FastNLS}, i.e., 
\begin{equation}\label{nls} \vspace{-4.5pt}
\hat{f}_0=\underset{f_0}{\arg\max} \ \underline{\mb{y}}_{\text{W}}^T\mathbf{Z}(f_0)\left[ \mathbf{Z}^H(f_0)\mathbf{Z}(f_0)\right] ^{-1}\mathbf{Z}^H(f_0)\underline{\mb{y}}_{\text{W}}
\end{equation} \normalsize
for all candidate model orders, including $L=0$ as a candidate to do voicing detection. The final model order $L$ is selected using model selection criteria such as Bayesian Information Criteria (BIC) \cite{bayesianStoica}. Here $\underline{\mb{y}}_{\text{W}}$ denotes the pre-whitened signal vector, where an underlined vector has $N$ (or even $N_{\text{opt}}$) samples.
\item \noindent The amplitude vector is estimated via least-squares as $\hat{\bm{\alpha}}=[\mathbf{Z}^H(\hat{f_0})\mathbf{Z}(\hat{f_0})]^{-1}\mathbf{Z}(\hat{f_0})^H\mathbf{y}$ \cite{MadsPitch}, after which the AR parameters of the residual $\underline{\mb{x}}=\underline{\mb{y}}-\mathbf{Z}(\hat{f_0})\hat{\bm{\alpha}}$ (and also $\hat{\mathbf{R}}_x$) are directly obtained \cite{stoica2005spectral}. These are directly used as the coefficients of a new AR pre-whitening filter, which is applied to yield $\underline{\mb{y}}_{\text{W}}$.
\end{enumerate}
The iterations are stopped when the difference of the cost function in (\ref{nls}) between two consecutive iterations is below a threshold value, or a maximum number of iterations is reached \cite{robust2020}. 
The estimation of the parameters for the different segment lengths allows us to obtain the segmentation markers for voiced speech extraction as described in the next section. Once these markers have been obtained, the noisy speech is processed to estimate the parameters inside the segments of 
length $N_{\text{opt}}$, from which $\mathbf v$ can be extracted using the matrix (\ref{voicFilter}).  

To obtain an estimate of the unvoiced part, we consider the modelled stochastic sequence  $\underline{\mb{x}}=\underline{\mb{y}}-\mathbf{Z}(\hat{f_0})\hat{\bm{\alpha}}$. The processing to estimate $u(n)$ is also obtained from an adaptive segmentation, but which is different from the one employed to extract the voiced part, i.e., the model in (\ref{unvModel}) will have a more accurate fit for an optimal segment length $N_{\text{opt}}' (\neq N_{\text{opt}})$. From pre-trained spectral shapes with the corresponding excitation variances, the modelled spectrum of the stochastic part is written as $\hat{\Phi}_x(\omega)=\frac{\sigma_u^2}{\left|B_u(\omega)\right|^2}+\frac{\sigma_c^2}{\left|\Gamma_c(\omega)\right|^2}$, where $\sigma_u^2$ and $\sigma_c^2$ are the excitation variances of unvoiced speech and noise, and $
B_u(\omega)=1+\sum_{i=1}^{P}\beta_{u_i}e^{-j\omega i}, \, \Gamma_c(\omega)=1+\sum_{i=1}^{P} \gamma_{c_i}e^{-j\omega i}$. The parameters to be estimated are $\{ \sigma_u^2,\sigma_c^2, \{\beta_{u_i}\}_{i=1}^P, \{\gamma_{c_i}\}_{i=1}^P \}$. Denoting $\beta_u^i(\omega)$ and $\gamma_c^j(\omega)$ the spectra of the $i^{th}$ and $j^{th}$
unvoiced speech and noise codebook entries, the single indices corresponding to the approximate ML estimate of the AR spectral shapes are obtained as \vspace{-5.5pt}
 \begin{equation} \label{minimIS} \scriptsize
    \{i^*,j^*\} = \arg \min_{i,j}\min_{\sigma_u^2,\sigma_c^2}d_{\text{IS}}(\hat{\Phi}_x,\frac{\sigma_u^2}{\left|B_u^i(\omega)\right|^2}+\frac{\sigma_c^2}{\left|\Gamma_c^j(\omega)\right|^2}), 
\end{equation} \normalsize
where $d_{\text{IS}}$ is the Itakura-Saito distance. For all combinations, the excitation variances are needed and are obtained as described in \cite{srinivasan2005codebook}. Similarly to the voiced case, in order to estimate $N_{\text{opt}}'$, we first need to estimate the parameters for all the possible segment lengths. The optimal length will maximise the log-likelihood function, as described later. Having obtained the optimal codebook entries and excitation variances on the segment of length $N_{\text{opt}}'$, they are used to form a Wiener filter \vspace{-5.5pt}
\begin{equation} \label{WienerFrequency} \small
H_u(\omega) = \frac{\frac{\sigma_u^{2*}}{\left|B_u^{i*}(\omega)\right|^2}}{\frac{\sigma_u^{2*}}{\left|B_u^{i*}(\omega)\right|^2}+\frac{\sigma_c^{2*}}{\left|\Gamma_c^{j*}(\omega)\right|^2}}, \normalsize
\end{equation}
which is applied to $\underline{\underline{\mathbf{x}}}$ (of length $N_{\text{opt}}'$) in order to extract $u(n)$. 

\section{Criteria for optimal segmentation} \vspace{-2.5pt}

Based on the principle of \cite{RDprediction}, in \cite{ColoredWhite} it was proposed to segment the signal based on the MAP criterion which assumes a WGN condition. To deal with colored noise, it is therefore required to pre-whiten $y(n)$ \cite{eusipco2019}. The segmentation markers are required before applying the linear filtering to extract the voiced part. Each way in which the signal can be segmented (i.e., a segment composed of a number of minimum-length segments) is considered as a model, among a set of candidate models $\mathcal{M}$. Under the MAP criterion, the model which maximimizes the model a posteriori probability given the observation, will be selected. The criterion \cite{MadsPitch, djuric6model,ColoredWhite} consists of a data log-likelihood term, and a term which penalizes model complexity. The estimated model order $\hat{L}$ is a function of $N$, in which $\hat{L}(N)$ and $\hat{f}_0(N)$ are estimated for each candidate segment with the iterative procedure described in \cite{robust2020}. For a candidate segment detected as voiced, i.e., $\hat{L}(N) \neq 0$, and considering the real signal harmonic model, the MAP cost function is
\begin{equation} \small\scriptsize \vspace{-5.5pt}
 J_1(N) = \frac{N}{2} \ln \frac{1}{N}  || \underline{\mb y}_{\text{W}} - \mb {Z \bm{\alpha_{\text{W}}}} ||_2^2 + \frac{3}{2} \ln N + \hat{L} (N) \ln N,
\label{eq:map_harmonic}
\end{equation} \normalsize 
in which the amplitude vector $\bm{\alpha}_{\text{W}}$ is obtained in this case from the pre-whitened signal. If a candidate segment is detected as not-voiced, i.e., $\hat{L}(N)=0$, the MAP cost function involved in the comparison is instead $J(N) = \frac{N}{2} \ln ||\underline{\mb{y}}_{\text{W}}||_2^2$. After the extraction of voiced speech, the modelled residual $x(n)$ is segmented based on the log-likelihood 
\begin{equation} \scriptsize \vspace{-6.5pt}
 J_2(N) = \frac{N}{2}d_{\text{IS}}(\hat{\Phi}_x,\frac{\sigma_u^2}{\left|B_u^i(\omega)\right|^2}+\frac{\sigma_c^2}{\left|\Gamma_c^j(\omega)\right|^2})+\frac{1}{2}\sum_{k=1}^{N}\ln \hat{\Phi}_x .
\label{eq:logLikelihood}
\end{equation} \normalsize 
The model which maximises the log-likelihood given the observed residual will be selected. The markers are required before applying the filter in (\ref{WienerFrequency}).

The segmentation requires that the cost is additive and independent over the segments, which is satisfied for both previous criteria. The optimal lengths $N_{\text{opt}}$ and $N_{\text{opt}}'$ are found by comparing the cost of all the possibilities from the set of segment lengths and choosing the one minimizing the cost over all candidates, i.e., $\widehat{\mathcal M} = \arg \min_{\mathcal M} J_i$, $i \in\{1,2\}$. A minimal segment length, $N_{\text{min}}$, is defined, generating a subsegment of $N_{\text{min}}$ samples and dividing the signal into $S$ subsegments. This gives $2^{S-1}$ ways of segmenting the signal into $S$ subsegments, and a maximum number of subsegments $B_{\text{max}}$ is set. A dynamic programming algorithm is then used to find the optimal number of subsegments in a segment, $b_{\text{opt}}$, for all subsegments, $s=1,...,S$, starting at $s = 1$ moving continuously to $s=S$ \cite{RDprediction}. For every subsegment, the cost of all new subsegment combinations are reused from earlier subsegments. When the end of the signal is reached, the optimal segmentation of the signal is found, starting at the last subsegment and jumping backwards through the signal until reaching the beginning. This is done by starting at $s = S$ and setting the number of subsegments in the last segment to $b_\text{opt}(S)$. Thereby, the next segment ends at subsegment $s = S-b_\text{opt}(M)$ and includes $b_\text{opt}(S-b_\text{opt}(S))$ subsegments. This is continued until $s=0$. The segmentation algorithm is described in \cite{ColoredWhite}. \vspace{-2pt}

To summarize, the steps to decompose (offline) noisy speech into its voiced and unvoiced components are: \vspace{-2pt}
\begin{enumerate} [wide, labelwidth=!, labelindent=0pt]
\item The noisy signal is pre-processed with an adaptive autoregressive pre-whitener \cite{eusipco2019}, yielding $y_{\text{W}}(n)$.  
\item Parameter estimates of $v(n)$ and $x(n)$ are jointly obtained \cite{robust2020} for all candidate segment lengths. Followingly, based on (\ref{eq:map_harmonic}), the markers of the optimal segmentation for voiced speech and $N_{\text{opt}}$ are obtained. 
\item Parameter estimates of $v(n)$ and $x(n)$ and statistics $\mathbf{R}_{\mathbf{v}}$, $\mathbf{R}_{\mathbf{x}}$ are obtained from the segments of length $N_\text{opt}$. If $\hat{L}(N_{\text{opt}}) \neq 0$, estimate $\mathbf{v}$ using (\ref{voicFilter}) after joint diagonalization of $\mathbf{R}_{\mathbf{v}}$ and $\mathbf{R}_{\mathbf{x}}$.  
\item Obtain the modelled residual $\underline{\mb{x}}=\underline{\mb{y}}-\mathbf{Z}(\hat{f_0})\hat{\bm{\alpha}}$ in all the different obtained optimal lengths $\{N_{\text{opt}}\}$. Once the whole modelled $x(n)$ is obtained, estimate unvoiced speech parameters $\{ \sigma_u^2, \{\beta_{u_i}\}_{i=1}^P\}$ for all candidate segment lengths. 
\item Based on (\ref{eq:logLikelihood}), obtain the markers of the optimal segmentation for unvoiced speech and $N_{\text{opt}}'$. 
\item The unvoiced speech parameters $\{ \sigma_u^2, \{\beta_{u_i}\}_{i=1}^P\}$ are obtained from the segments of length $N_{\text{opt}}'$. Extract $\underline{\underline{\mathbf{u}}}$ using (\ref{WienerFrequency}).
\end{enumerate}

\section{Experimental evaluation}

\begin{figure}[t]
	\centering
	{\includegraphics[width=0.48\textwidth, height=0.27\textheight]{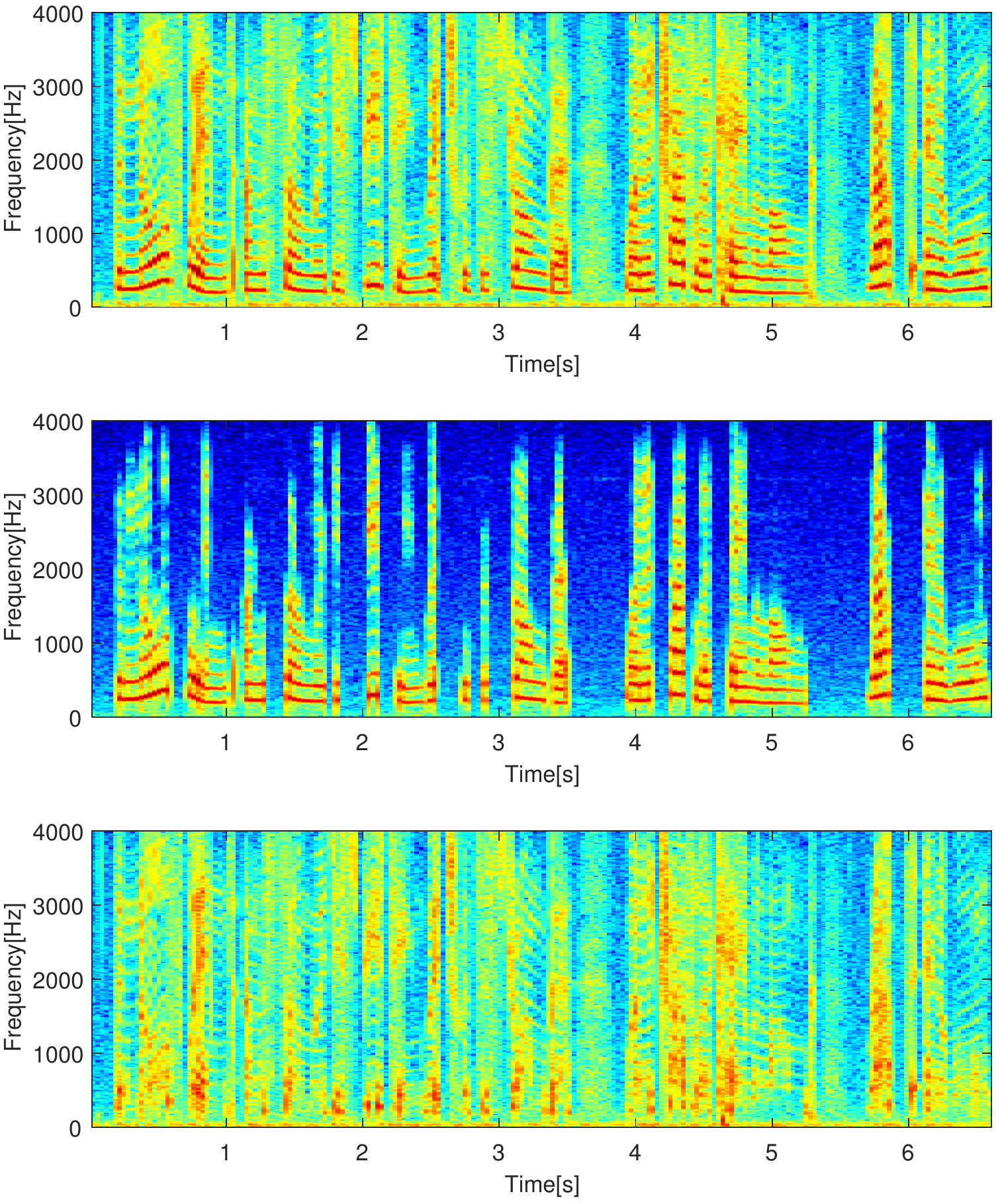}}
	\caption[caption]{From top to bottom: Spectrograms of the observed signal,  extracted voiced part, and extracted unvoiced part.}
	\label{Figure1}
\end{figure}\vspace{-0.10cm}

\begin{figure}[b]
	\centering
	{\includegraphics[width=0.43\textwidth, height=0.20\textheight]{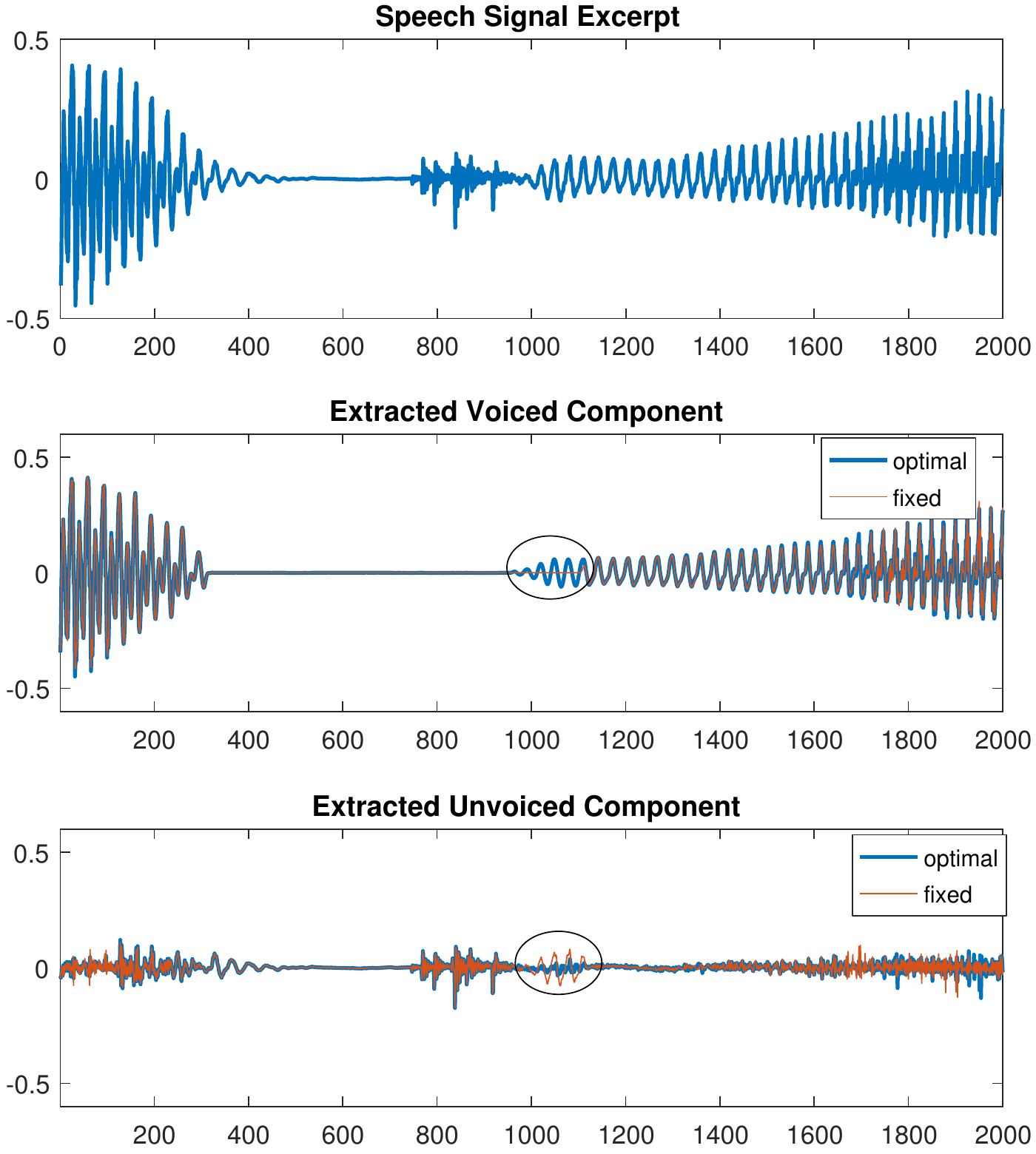}}
	\caption[caption]{Extraction of voiced and unvoiced components from optimal and fixed segmentation on a clean signal excerpt. }
	\label{Figure2}
\end{figure}\vspace{-0.13cm}

We first illustrate the extracted speech components of one of the clean female excerpts from the Keele database \cite{keeleref}, after the voiced speech segmentation markers were obtained. Although subsegments increasing in steps of 5 ms (i.e., $N =40$ at 8 kHz) are considered for $v(n)$, only segments from 20 to 50 ms (i.e. $N=160$ to $N=400$ in steps of 40) are possible in the segmentation. That is, the maximum number of possible subsegments is $B_{\text{max}}=10$, and the cost for $b=1$, $b=2$ or $b=3$ is set to infinity, as the pitch estimator does not work well for very short segment lengths and low pitch $f_0$. For the optimal segments for which $\hat{L}(N) \neq 0$, the filtering matrix (\ref{voicFilter}) with $M=40$ is applied, and the filtering is updated every 20 samples, i.e., there is a 50 \% of overlap. $M=40$ was chosen as it is an integer divisor of all the candidate segment lengths, which facilitates the processing. The difference from the clean signal and $v(n)$ corresponds to $u(n)$, and this corresponds to a ground truth for the unvoiced speech component. The spectrograms of $s(n)$ and its corresponding $v(n)$ and $u(n)$ are displayed on Fig.1. It is seen that $v(n)$ has an appearance with horizontal striations and that $u(n)$ is displayed by rectangular patterns over a wide range of frequencies. Around 2.1 s, the harmonics up to around 3 kHz are obtained in $v(n)$. An example of how the time series of $v(n)$ and $u(n)$ look like if either a segment of fixed length is used (here 20 ms), or if the extraction of $v(n)$ is done using the optimal segmentation, is displayed in Fig. 2. The unvoiced part obtained from the optimal segmentation used for $v(n)$ exhibits a more stochastic nature compared to the one obtained from using segments of a fixed size to extract $v(n)$. The marked region exhibits a periodic nature, which corresponds to $v(n)$. The optimal segmentation results in a better modelling of the periodic parts in the extracted voiced component. We now proceed to evaluate the decomposition performance in noisy conditions. Four excerpts of 4 s of the Keele database files were added babble, factory, street and restaurant noise, at iSNRs of 0, 5 and 10 dB. The performance per iSNR is presented averaged across all the noise types, and two runs are done per excerpt and noise type. That is, a total of 32 runs are considered per iSNR. Before applying the segmentation, the signal was pre-whitened from the setup described in \cite{eusipco2019}, which relies on a parametric NMF noise PSD estimate. The codebook of AR entries of unvoiced speech (including also from silent segments) was obtained from the training on samples which correspond to the difference of clean signals and the voiced speech extracted from the Wiener filter. And as stated before, this corresponds to a ground truth of unvoiced speech. The samples used for training were different than those at evaluation. The training was done on segments of length $N=160$ (i.e., 20 ms) with an overlap of 50 \% between them, with an AR order $P = 14$. Similarly, the codebook of AR entries of noise (including babble, F-16, restaurant and factory \cite{varga1993assessment}) was trained. A total of 64 unvoiced speech and 16 noise entries were obtained from a standard vector quantization technique \cite{linde1980algorithm}. When evaluating (\ref{minimIS}), the modelled  $\underline{\bm x}$ was fitted to an AR spectrum $\Phi_x$ of order 28  \cite{srinivasan2005codebook}. To extract $u(n)$, segments from 15 to 40 ms were made possible for the segmentation. The results are shown in Figure 3, comparing the performance of applying the optimal segmentation to the extraction based on a traditional fixed one (20 ms). The decomposition performance is evaluated in terms of segSNR and Log Spectral distance (LSD) \cite{srinivasan2005codebook}. It is also compared to the decomposition methods \cite{elie2016robust, Yegnanarayana} after OMLSA speech enhancement \cite{cohen2002optimal} was applied as a pre-processor. This is done to attenuate the noise which is not taken into account in them. The comparison also has the case where noisy speech is obtained as an estimate, in order to see if the methods  perform better than the case of not processing the signal. At an iSNR of 10 dB, the extraction based on adaptive segments leads to a higher segSNR for the case of $v(n)$. Also, with respect to the LSD, lower values are obtained for both the extracted $v(n)$ and $u(n)$ based on adaptive segments. At an iSNR of 5 dB, although the confidence intervals of segSNR overlap, as seen from the extreme intervals, there is higher probability that the adaptive segmentation leads to a better recovery of the components, and also the LSD values are clearly separated. At 0 dB, both ways of segmenting lead to similar performance. From using optimal segmentation, it is possible to get lower LSD for $v(n)$ compared to \cite{elie2016robust}, which based its processing on segments of fixed size. Although it is possible to achieve higher segSNR with the proposed, it is seen that the other methods combined with enhancement achieve lower LSD for $u(n)$, at lower SNRs. However, there is a potential to trade off distortion and noise reduction by considering other filters in the VSLF framework \cite{benesty2016signal}.  

\begin{figure}[t]
	\centering
	{\includegraphics[width=0.48\textwidth, height=0.28\textheight]{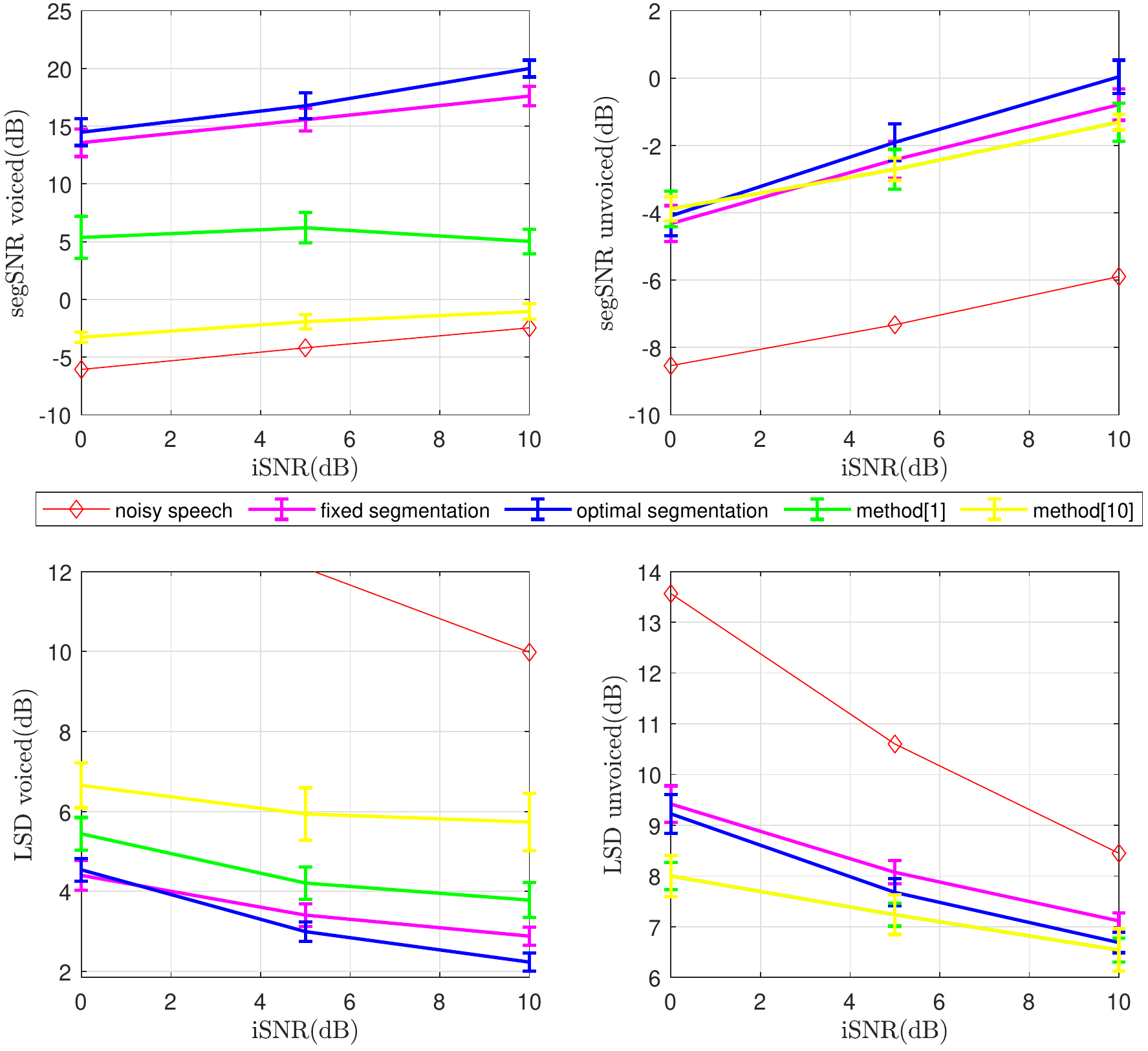}}
	\vspace{-0.85pt}
	\caption[caption]{LSD and segmental SNR (segSNR) in different iSNRs averaged across four noise types. }\vspace{-0.25cm}
	\label{Figure3}
\end{figure}\vspace{-0.34cm}

\section{Discussion}\vspace{-4pt}

The use of an optimal segmentation combined with parameter estimates of an hybrid speech model allow to have a more accurate recovery of the voiced and unvoiced speech parts, compared to the use of fixed segments. Specifically, an adaptive segmentation results in a better modelling of the periodic parts in the voiced component with a higher probability of improved segSNR and also of a lower LSD of both extracted voiced and unvoiced parts. We considered prior spectral information stored in codebooks in order to differentiate between unvoiced speech and noise. A higher segSNR and lower LSD for the voiced part is possible when compared to reference methods, with a potential to reduce the LSD for the extracted unvoiced part. As future work, we will consider deriving the segmentation based on the recently introduced joint pitch-AR estimator \cite{QUINN2021107860}. 

\atColsBreak{\vskip5pt}
\bibliographystyle{IEEEtran}

\bibliography{mybib}

\begin{thebibliography}{10}
\providecommand{\url}[1]{#1}
\csname url@samestyle\endcsname
\providecommand{\newblock}{\relax}
\providecommand{\bibinfo}[2]{#2}
\providecommand{\BIBentrySTDinterwordspacing}{\spaceskip=0pt\relax}
\providecommand{\BIBentryALTinterwordstretchfactor}{4}
\providecommand{\BIBentryALTinterwordspacing}{\spaceskip=\fontdimen2\font plus
\BIBentryALTinterwordstretchfactor\fontdimen3\font minus
  \fontdimen4\font\relax}
\providecommand{\BIBforeignlanguage}[2]{{%
\expandafter\ifx\csname l@#1\endcsname\relax
\typeout{** WARNING: IEEEtran.bst: No hyphenation pattern has been}%
\typeout{** loaded for the language `#1'. Using the pattern for}%
\typeout{** the default language instead.}%
\else
\language=\csname l@#1\endcsname
\fi
#2}}
\providecommand{\BIBdecl}{\relax}
\BIBdecl

\bibitem{elie2016robust}
B.~Elie and G.~Chardon, ``Robust tonal and noise separation in presence of
  colored noise, and application to voiced fricatives,'' in \emph{22nd
  International Congress on Acoustics (ICA)}, 2016.

\bibitem{griffin1988multiband}
D.~W. Griffin and J.~S. Lim, ``Multiband excitation vocoder,'' \emph{IEEE
  Transactions on acoustics, speech, and signal processing}, vol.~36, no.~8,
  pp. 1223--1235, 1988.

\bibitem{aej2018}
A.~E. {Jaramillo}, J.~K. {Nielsen}, and M.~G. {Christensen}, ``On optimal
  filtering for speech decomposition,'' in \emph{26th European Signal
  Processing Conference (EUSIPCO)}, 2018.

\bibitem{mehta2005synthesis}
D.~Mehta and T.~F. Quatieri, ``Synthesis, analysis, and pitch modification of
  the breathy vowel,'' in \emph{IEEE Workshop on Applications of Signal
  Processing to Audio and Acoustics, 2005.}\hskip 1em plus 0.5em minus
  0.4em\relax IEEE, 2005, pp. 199--202.

\bibitem{jackson2000characterisation}
P.~J. Jackson, ``Characterisation of plosive, fricative and aspiration
  components in speech production,'' Ph.D. dissertation, University of
  Southampton, 2000.

\bibitem{stylianou1995high}
Y.~Stylianou, J.~Laroche, and E.~Moulines, ``High-quality speech modification
  based on a harmonic+ noise model,'' in \emph{Fourth European Conference on
  Speech Communication and Technology}, 1995.

\bibitem{belalcazar2013automatic}
E.~Belalcazar-Bolanos, J.~Orozco-Arroyave, J.~Arias-Londono, J.~Vargas-Bonilla,
  and E.~N{\"o}th, ``Automatic detection of {P}arkinson's disease using noise
  measures of speech,'' in \emph{Symposium of Signals, Images and Artificial
  Vision-2013: STSIVA-2013}.\hskip 1em plus 0.5em minus 0.4em\relax IEEE, 2013,
  pp. 1--5.

\bibitem{stylianou2004modeling}
Y.~Stylianou, ``Modeling speech based on harmonic plus noise models,'' in
  \emph{International School on Neural Networks, Initiated by IIASS and
  EMFCSC}.\hskip 1em plus 0.5em minus 0.4em\relax Springer, 2004, pp. 244--260.

\bibitem{jensen2001speech}
J.~Jensen and J.~H. Hansen, ``Speech enhancement using a constrained iterative
  sinusoidal model,'' \emph{IEEE Transactions on Speech and Audio Processing},
  vol.~9, no.~7, pp. 731--740, 2001.

\bibitem{MadsPitch}
M.~G. Christensen and A.~Jakobsson, \emph{Multi-Pitch Estimation}, ser.
  Synthesis Lectures on Speech and Audio Processing.\hskip 1em plus 0.5em minus
  0.4em\relax Morgan {\&} Claypool Publishers, 2009.

\bibitem{FastNLS}
J.~K. Nielsen, T.~L. Jensen, J.~R. Jensen, M.~G. Christensen, and S.~H. Jensen,
  ``Fast fundamental frequency estimation: Making a statistically efficient
  estimator computationally efficient,'' \emph{Signal Processing}, vol. 135,
  no. Supp. C, pp. 188 -- 197, 2017.

\bibitem{QUINN2021107860}
B.~G. Quinn, J.~K. Nielsen, and M.~G. Christensen, ``Fast algorithms for
  fundamental frequency estimation in autoregressive noise,'' \emph{Signal
  Processing}, vol. 180, p. 107860, 2021.

\bibitem{Yegnanarayana}
B.~{Yegnanarayana}, C.~{d'Alessandro}, and V.~{Darsinos}, ``An iterative
  algorithm for decomposition of speech signals into periodic and aperiodic
  components,'' \emph{IEEE Transactions on Speech and Audio Processing},
  vol.~6, no.~1, pp. 1--11, 1998.

\bibitem{moran2006telephony}
R.~J. Moran, R.~B. Reilly, P.~de~Chazal, and P.~D. Lacy, ``Telephony-based
  voice pathology assessment using automated speech analysis,'' \emph{IEEE
  Transactions on Biomedical Engineering}, vol.~53, no.~3, pp. 468--477, 2006.

\bibitem{robust2020}
A.~E. {Jaramillo}, A.~{Jakobsson}, J.~K. {Nielsen}, and M.~{G. Christensen},
  ``Robust fundamental frequency estimation in coloured noise,'' in \emph{IEEE
  International Conference on Acoustics, Speech and Signal Processing
  (ICASSP)}, 2020, pp. 741--745.

\bibitem{preference}
K.~K. {Paliwal}, J.~G. {Lyons}, and K.~K. {Wójcicki}, ``Preference for 20-40
  ms window duration in speech analysis,'' in \emph{4th International
  Conference on Signal Processing and Communication Systems}, 2010, pp. 1--4.

\bibitem{drepper2005two}
F.~R. Drepper, ``A two-level drive--response model of non-stationary speech
  signals,'' in \emph{International Conference on Nonlinear Analyses and
  Algorithms for Speech Processing}.\hskip 1em plus 0.5em minus 0.4em\relax
  Springer, 2005, pp. 125--138.

\bibitem{ColoredWhite}
S.~M. Nørholm, J.~R. Jensen, and M.~G. Christensen, ``Instantaneous
  fundamental frequency estimation with optimal segmentation for nonstationary
  voiced speech,'' \emph{IEEE/ACM Transactions on Audio, Speech, and Language
  Processing}, vol.~24, no.~12, pp. 2354--2367, Dec 2016.

\bibitem{Martin2001}
R.~Martin, ``Noise power spectral density estimation based on optimal smoothing
  and minimum statistics,'' \emph{IEEE Transactions on Speech and Audio
  Processing}, vol.~9, no.~5, pp. 504--512, Jul. 2001.

\bibitem{srinivasan2005codebook}
S.~Srinivasan, J.~Samuelsson, and W.~B. Kleijn, ``Codebook driven short-term
  predictor parameter estimation for speech enhancement,'' \emph{IEEE
  Transactions on Audio, Speech, and Language Processing}, vol.~14, no.~1, pp.
  163--176, 2005.

\bibitem{eusipco2019}
A.~{E. Jaramillo}, J.~{K. Nielsen}, and M.~{G. Christensen}, ``Adaptive
  pre-whitening based on parametric {NMF},'' in \emph{2019 27th European Signal
  Processing Conference (EUSIPCO)}, September 2019.

\bibitem{RDprediction}
P.~{Prandoni} and M.~{Vetterli}, ``R/{D} optimal linear prediction,''
  \emph{IEEE Transactions on Speech and Audio Processing}, vol.~8, no.~6, pp.
  646--655, 2000.

\bibitem{norholm2015enhancement}
S.~M. N{\o}rholm, \emph{Enhancement of speech signals-with a focus on voiced
  speech models}, ser. Ph.D. thesis, Aalborg Universitet, 2015.

\bibitem{benesty2016signal}
J.~Benesty, M.~G. Christensen, and J.~R. Jensen, \emph{Signal enhancement with
  variable span linear filters}.\hskip 1em plus 0.5em minus 0.4em\relax
  Springer, 2016, vol.~7.

\bibitem{stoica2005spectral}
P.~Stoica and R.~L. Moses, ``Spectral analysis of signals,'' \emph{Pearson},
  2005.

\bibitem{bayesianStoica}
P.~{Stoica} and Y.~{Selen}, ``Model-order selection: a review of information
  criterion rules,'' \emph{IEEE Signal Processing Magazine}, vol.~21, no.~4,
  pp. 36--47, July 2004.

\bibitem{djuric6model}
P.~M. Djuric, ``A model selection rule for sinusoids in white gaussian noise,''
  \emph{IEEE Transactions on Signal Processing}, vol.~44, no.~7, pp.
  1744--1751, 1996.

\bibitem{keeleref}
F.~Plante, G.~F. Meyer, and W.~A. Ainsworth, ``A pitch extraction reference
  database,'' in \emph{EUROSPEECH}, 1995.

\bibitem{varga1993assessment}
A.~Varga and H.~J. Steeneken, ``Assessment for automatic speech recognition:
  Ii. {NOISEX}-92: A database and an experiment to study the effect of additive
  noise on speech recognition systems,'' \emph{Speech communication}, vol.~12,
  no.~3, pp. 247--251, 1993.

\bibitem{linde1980algorithm}
Y.~Linde, A.~Buzo, and R.~Gray, ``An algorithm for vector quantizer design,''
  \emph{IEEE Transactions on communications}, vol.~28, no.~1, pp. 84--95, 1980.

\bibitem{cohen2002optimal}
I.~Cohen, ``Optimal speech enhancement under signal presence uncertainty using
  log-spectral amplitude estimator,'' \emph{IEEE Signal processing letters},
  vol.~9, no.~4, pp. 113--116, 2002.

\end{thebibliography}


\end{document}